\documentclass[twocolumn,showpacs,preprintnumbers,amsmath,amssymb]{revtex4}

\usepackage{graphicx}% Include figure files
\usepackage{dcolumn}% Align table columns on decimal point
\usepackage{bm}% bold math

\usepackage{color}

\newcommand{\beq}[1]{  \begin{equation} \label{#1} }  
\newcommand{\eeq}{     \end{equation}}  
\newcommand{\bal}[1]{\begin{align} \label{#1} }

\newcommand{\rf}[1]{(\ref{#1})}

\def\bd#1{\mbox{\boldmath$\displaystyle\mathbf{#1}$} }
\def\eps{\epsilon}

\begin{document} %%%%%%%%%%%%%%%%%%%%%%%%%%%%%%%%%%%%%%%%%%%%%%%%%%%%%%%%%%%%%%%%%%%%%%%%%

\title{Comment on ``Method to analyze electromechanical stability of dielectric elastomers" [Appl. Phys. Lett. 91, 061921 (2007)]}  

\author{ Andrew N. Norris} 

  \email{norris@rutgers.edu}
\affiliation{Mechanical and Aerospace Engineering, Rutgers University, Piscataway NJ 08854}

\date{\today}

\begin{abstract}

The model of Zhao and Suo can be readily generalized to predict  
 the critical breakdown electric field $E_c$ value of elastomers with arbitrary elastic strain energy function.   An explicit expression for $E_c$  is presented for elastomeric thin films under biaxial strain and comparisons are made with experimental data using a two term Ogden rubber elasticity model.  Simplified results for uniaxial and for equi-biaxial stress provide further insight into the findings of Zhao and Suo.

\end{abstract}


\maketitle

%\cite{Kleinert06}  consider the large-stiffness expansion of the partition function.

%\section{Introduction}

The paper of Zhao and Suo \cite{Zhao07} describes a fully nonlinear electromechanical  model for  the phenomenon of electrical breakdown in thin elastomers.   The purpose of this comment is to point out some analytical simplifications which provide further insight into their model, and to provide explicit formulas useful for elastomer design.  

The results here stem from the observation that the determinant of the Hessian $\bd H$ of eq. (4) in \cite{Zhao07} may be factored, leading to semi-explicit formulas for the critical values of the electrical and mechanical parameters.   
It  may be checked that the determinant reduces to  a quadratic in $z$,
\bal{1}
%\beq{1}
\det {\bd H} =& \frac{\mu^2 \epsilon^{-1}}{\lambda_1^8\lambda_2^8 }
\bigg[5+3(\lambda_1^2+\lambda_2^2) \lambda_1^2\lambda_2^2 +\lambda_1^6\lambda_2^6
\nonumber \\ \qquad & \qquad  
+[2-(\lambda_1^2+\lambda_2^2) \lambda_1^2\lambda_2^2]z - 3 z^2\bigg],
%\eeq
\end{align}
where the nondimensional parameter $z$ is
\beq{2}
z = (\mu \epsilon)^{-1}\widetilde{D}^2 
 = \mu^{-1}\epsilon \widetilde{E}^2  \lambda_1^4\lambda_2^4
  = \mu^{-1} \epsilon {E}^2  \lambda_1^2\lambda_2^2 , 
 % z = \frac{\widetilde{D}^2}{\mu \epsilon}  = \frac{\epsilon \widetilde{E}^2}{\mu } \lambda_1^4\lambda_2^4   = \frac{\epsilon {E}^2}{\mu } \lambda_1^2\lambda_2^2 , 
\eeq 
 and all other notation follows \cite{Zhao07}.
The roots of the quadratic  are real and of opposite sign, so  
 there is a unique  positive value of $z$ at which the Hessian  is no longer positive definite.  It turns out that the same 
 structure of the Hessian is retained for  free energy of the form 
\beq{0}
W (\lambda_1, \lambda_2,\widetilde{D})  = U(\lambda_1, \lambda_2) 
+ \frac{\widetilde{D}^2}{2 \epsilon}  \lambda_1^{-2}\lambda_2^{-2} , 
%+ \frac12 \epsilon E^2.
\eeq 
where $U(\lambda_1, \lambda_2) = \psi( \lambda_1, \lambda_2,
\lambda_1^{-1}\lambda_2^{-1})$. 
An equation similar to \rf{1} is obtained, and taking the single positive root shows that the  critical value of the electric field satisfies
 \begin{widetext}
\beq{-1} 
\epsilon E_c^2= \frac16\bigg(
4\lambda_1\lambda_2  U_{12} - \lambda_1^2 U_{11} -\lambda_2^2 U_{22}  
+ \sqrt{ (\lambda_1^2 U_{11} +\lambda_2^2 U_{22} - 4\lambda_1\lambda_2  U_{12} )^2
+ 12 \lambda_1^2\lambda_2^2( U_{11}U_{22}- U_{12}^2)}\bigg), 
\eeq
 \end{widetext}
where $U_{ij}= \partial^2 U/\partial\lambda_i\partial\lambda_j$.  
If the  stretches $\lambda_1$ and $\lambda_2$ are prescribed, then eq. \rf{-1} is sufficient to estimate the critical field strength.  Otherwise, if the  nominal stresses $s_1$ and $s_2$ are prescribed then the stretches are determined from 
\beq{5}
s_j= U_j  - \lambda_j^{-1 }
 \epsilon E_c^2 ,  \qquad
j=1,2. 
\eeq

Under  equi-biaxial strain $\lambda_1=\lambda_2=\lambda$, eq. \rf{-1} becomes
\beq{-16}
\epsilon E_c^2=  \frac{ \lambda^2}{3} \big( U_{11} +U_{12} \big)  , 
\eeq
where the critical value of the stress $s_1=s_2=s$ is
\beq{-57}
  U_1 - \frac{1}{3\lambda}( U_{11} +U_{12}) = s. 
\eeq
Consider the Ogden model for rubber \cite{Ogden84} 
\beq{044}
\psi( \lambda_1, \lambda_2,\lambda_3) = \sum_{p=1}^N \frac{\mu_p}{\alpha_p} \big(
\lambda_1^{\alpha_p}+\lambda_2^{\alpha_p}+\lambda_3^{\alpha_p}
\big), 
\eeq
for which the  critical electrical field strength is  
\beq{=16}
\epsilon E_c^2= \frac13 \sum_{p=1}^N \mu_p \big[
(\alpha_p -1)\lambda^{\alpha_p}+(2\alpha_p +1)\lambda^{-2\alpha_p}
\big]. 
\eeq
If the stress is prescribed then  $\lambda$ is given by 
\beq{+16}
\frac{1}{3\lambda}\sum_{p=1}^N  \mu_p\big[
(4- \alpha_p)\lambda^{\alpha_p}-(4+2\alpha_p )\lambda^{-2\alpha_p}
\big] = s.
\eeq
These  parameterize the critical electrical and mechanical fields in terms of $\lambda$.

%%%%%%%%%%%%%%%%%%%%%%%%%%%%%%%%%%%%%%%%%%%%%%%%%% Figure
\begin{figure}[htbp]
				\begin{center}	
				    \includegraphics[width=3.3in , height=2.6in 					]{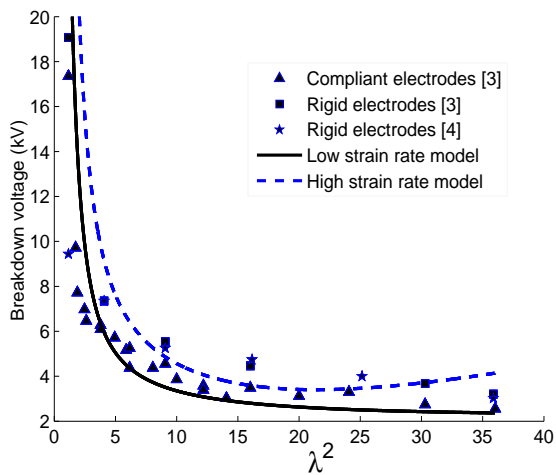} 
	\caption{The data show reported critical breakdown voltages as a function of the equi-biaxial  prestrain $\lambda$ for films of  VHB 4905/4910 elastomer, from \cite{Plante06,Kofod01}.  The curves are the predictions of eq. \rf{=16} using the elasticity parameters from \cite{Plante06}
	with $\eps_d = 12$. 
	    }
		\label{f1} \end{center}  
	\end{figure}
%%%%%%%%%%%%%%%%%%%%%%%%%%%%%%%%%%%%%%%%%%%%%%%%%% 

Values of the critical breakdown voltage for the elastomer VHB 4905/4910 have been reported by Plante and Dubowsky \cite{Plante06} and by Kofod et al. 
\cite{Kofod01}.  
Assuming the Ogden model with  $N=2$,  Plante and Dubowsky \cite{Plante06} measured values of 
$\alpha_1 = 1.445 (1.450)$, $\alpha_2 =  4.248 (8.360)$, $\mu_1 = 43,560(112,200)$ Pa,  $\mu_2 = 117.4(0.1045)$ Pa  for elastomer films of  initial thickness  $L_3= 1.5$ mm  at low (high) stretch rates.    
%The critical value  $\lambda_c$ from \rf{+16} is $1.2577$, which is  essentially unchanged whether  $N=1$ or $N=2$.  We find $\epsilon E_c^2=3.85\times 10^{4}$. 
Using these values the critical breakdown voltage $V_c=L_3\widetilde{E}_c $ predicted by eq.  \rf{=16} is compared with the data of 
\cite{Plante06,Kofod01} in Figure 1. 
The  material dielectric constant  was chosen as 
$\epsilon_d =12 $ to fit the curves with the data, where 
 $\epsilon =\epsilon_d \epsilon_0$ and  $\epsilon_0= 8.85\times 10^{-12}$ F/m is the free space permittivity.    
The agreement is reasonable, given that the experiments were not performed in a state of pure equi-biaxial  stress. 

Some useful explicit results can be determined for the one term  Ogden model
$ (N=1, \alpha_1, \mu_1, \rightarrow \alpha, \mu)$.   Under \emph{equi-biaxial} stress the critical stretch satisfies 
$\lambda \ge \lambda_c $ where $\lambda_c = \big( ({4+2\alpha})/({4-\alpha})\big)^{1/(3\alpha)}$ is the $s=0$ value.  This obviously  requires that $\alpha < 4$. 
The critical field 
$E_c$  has a unique minimum at $\lambda_0 =  \big(2 ({2\alpha +1})/({\alpha -1})\big)^{1/(3\alpha)}$ if $\alpha >1$.   
Zhao and Suo \cite{Zhao07} considered $\alpha = 2$, for which 
$\lambda_c  \approx 1.26$,  $\lambda_0 
 \approx 1.47$ and the minimum value of 
$\sqrt{\frac{\epsilon}{\mu }}  E_c$ is $1.038$.  

Finally, we note that the neo-Hookean constitutive model of Zhao and Suo \cite{Zhao07} is apparently unique among the $N=1$ Ogden models in that it yields a simple formula for  
 \emph{uniaxial stress}.  Thus,  eq. \rf{5} with $N=1$, $\alpha=2$ for $j=2$ and $s_2=0$ yields the  relation $\lambda_1^2  =  3\lambda_2^2/(\lambda_2^6 -1)$ between the stretches. 
Hence, we can parameterize the critical values in terms of 
$1 < \lambda_2 \le \lambda_c \approx 1.26$: 
\begin{subequations}\label{7}  \bal{7a}
%\beq{7}
\sqrt{\frac{\epsilon}{\mu }}  E_c
= & \big( \frac23 \lambda_2^2  +\frac13 \lambda_2^{-4}  \big)^{1/2},
%\lambda_1^2 \big(\lambda_1^2 - \lambda_1\frac{s_1}{\mu}\big)^2 -1,
 \\
 \frac{s_1}{\mu} =& \frac{\lambda_2}{\sqrt{3}} 
  \frac{(4-\lambda_2^6
  )}{\sqrt{\lambda_2^6 - 1}} .
%\lambda_2 &= \sqrt{\lambda_1^2 - \lambda_1\frac{s_1}{\mu}}, 
%\eeq
\label{7b} \end{align} \end{subequations}
In this case $E_c$ is a monotonically decreasing function of the stress $s_1$, and 
$\sqrt{\frac{\epsilon}{\mu }}  E_c \rightarrow 1$ in the limit of large uniaxial stress.   Figure 3(b) in \cite{Zhao07} indicates that this is the smallest achievable value of the critical electric field strength.   Generalization of the formulas \rf{7} to $\alpha \ne 2$ is possible but far more complicated.

%This reduces to eq. \rf{5} for the neo-Hookean strain energy considered in \cite{Zhao07}. 
%$U(\lambda_1, \lambda_2, \lambda_3)= \frac{\mu}{2}( \lambda_1^2+ \lambda_2^2+ \lambda_3^2 - 3)$

In summary, the model of Zhao and Suo readily generalizes to arbitrary elastic strain energy.   The explicit results reported here, such as 
eq.  \rf{-1},  can be used to compare different elastic constitutive models, and should be helpful in the design of elastomeric actuators.

\end{document}